# A Cluster-based Approach for Outlier Detection in Dynamic Data Streams (KORM: k-median OutlieR Miner)

Parneeta Dhaliwal , MPS Bhatia and Priti Bansal

**ABSTRACT**—Outlier detection in data streams has gained wide importance presently due to the increasing cases of fraud in various applications of data streams .The techniques for outlier detection have been divided into either statistics based , distance based , density based or deviation based. Till now, most of the work in the field of fraud detection was distance based but it is incompetent from computational point of view. In this paper we introduced a new clustering based approach, which divides the stream in chunks and clusters each chunk using k-median into variable number of clusters. Instead of storing complete data stream chunk in memory, we replace it with the weighted medians found after mining a data stream chunk and pass that information along with the newly arrived data chunk to the next phase. The weighted medians found in each phase are tested for outlierness and after a given number of phases, it is either declared as a real outlier or an inlier. Our technique is theoretically better than the k-means as it does not fix the number of clusters to k rather gives a range to it and provides a more stable and better solution which runs in poly-logarithmic space.
.

—————————— ◆ ——————————

## 1 INTRODUCTION

Data stream is an ordered sequence of unbounded objects $X_1$ ,$X_2$, $X_3$,…, $X_m$ where data arrives continuously , and can be read only once or a small number of times and it is practically infeasible to store it completely in memory. Each reading of the stream chunk is called a linear scan or a phase.

The increasing applications of data streams in the fields of fraud detection, network flow monitoring and data communications, has led to an increasing demand for data stream mining. Anomaly detection deals with the detection of data elements that are different from all the other elements in the data set. Traditional methods for anomaly detection dealt with data sets that were static and the data could be accessed a number of times so these methods cannot give efficient results for the analysis of data streams.

A lot of work for anomaly detection in data streams has been done traditionally by Knorr and Ng [8] using distance based approach. These techniques were highly dependent on the parameters provided by the users and were computationally expensive when applied to unbounded data streams. Later LOF [3] and its extension [11] were developed but the computation of LOF values required large number of nearest neighbour searches.

The evolution of data streams led to the change in the characteristics of the data streams like dimensionality, object features so a point which is an outlier in one phase may become an inlier in the next phase.

―――――――――――――――
- *Parneeta Dhaliwal is working as Senior Lecturer with KIIT College of Engineering, Gurgaon*
- *MPS Bhatia is working with as Professor in Netaji Subhas Institute of Technology, University of Delhi, New Delhi.*
- *Priti Bansal is working as Senior Lecturer Computer science with Ideal Institute of Technology, Ghaziabad*

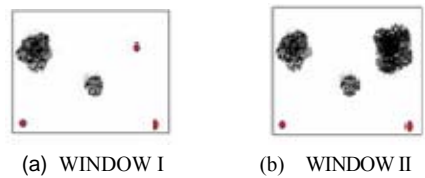

(a) WINDOW I   (b) WINDOW II

Fig. 1. Evolution of data stream from (a) to (b)

Fig. 1 shows two windows. three points appearing as temporal outliers in WINDOW I are left with only two outliers in WINDOW II, due to dynamics of the data streams. One of the outliers in WINDOW I, has a dense region around it in WINDOW II making it an inlier. Anguilli [5] developed two algorithms based on k-nearest neighbor approach but it required two parameters R and k from the user and involved each time pair-wise distance calculations which was very time consuming .

To overcome the above problems of pair-wise distance calculations and to let user free of providing sensitive parameters and to mine data streams in a resource constrained environment , Manzoor and Li[ 4] , proposed a clustering based approach for outlier detection based on k-mean . Here the data stream was divided into chunks of data and mined for temporal outliers, which over a number of phases were tested for being a final outlier or an inlier.

To overcome the limitation of fixed number of clusters k as in k-mean algorithm for outlier detection [4], we have proposed a k-median algorithm that gives us a range of k-values in the intermediate steps that provides more stability to our algorithm than the k-means [4]. The k-Mean objective function is to minimize the largest assignment distance, whereas in k-Median the sum of assignment distances (replaced by their squares) and the total cost is to be minimized.

The k-Mean measure is more sensitive to outliers than the k-Median objective so the k-median also gives us better clustering solution as well as better defined outliers. Hence, our approach is supposed to be better than the so k-mean [4] approach as the number of centers is unrestricted, but there is an additional cost for each center included in the solution.

In our algorithm we have tried to use the algorithm discussed in [1], which is a k-median algorithm for clustering data streams that results in a constant factor approximation in one pass using storage space O(k poly log n). Here, the data stream is divided into chunks of data which is mined in phases. In each phase, we find out the weighted medians (facilities). The data that has been read is deleted from memory and is replaced by the weighted medians. The facilities that have no increase in weight are considered as temporal candidate outliers for that phase. The next $phase_{i+1}$ takes the data that has not been read along with the weighted medians found till $phase_i$. The temporal outliers are checked for its outlierness for a given number of phases after which it is either declared as an inlier or a real outlier. After it has been declared as a real outlier it is further never used for clustering. The total cost of the product is the sum of all the facility costs(depending on the number of medians) and the service costs (cost of assigning a point to an already opened closest facility).

Section 2 deals with the related work, section 3 contains the problem definition, section 4 contains the proposed algorithm in detail and and section 5 gives the algorithm description, section 6 gives the experimental evaluation and section 7 gives the summary of the paper and the future work.

## 2 RELATED WORK

Most of the work on outlier detection was based on distance based approach. Knorr and Ng [8] proposed a distance based approach based on two parameters R and k. A data point is an outlier if less than k points lie within distance R from it. Ramaswamy et al[2] extended this idea further by giving a rank to the outliers. Given k and n, a data point is an outlier if no more than n-1 other points have higher value for $D_k$ than o, where $D_k$ denotes the distance of the $k^{th}$ nearest neighbor of o. Distance based outlier detection algorithms are unsuitable for clusters with varying densities and have large computational complexity. They are designed to work under the assumption that whole data set is stored in secondary memory and so they are not suitable for data streams. Breunig [3] introduced the notion of the local outlier factor LOF (a density based approach), which captures relative degree of isolation. Outlierness here is not considered a binary property, but a point is treated as an outlier according to its degree of isolation from its surrounding neighborhood.

Arning [9] described a deviation based technique for outlier detection where a point is considered an outlier if its features deviates from the features of other data points. For finding outliers in high dimensional data, Aggarwal [6] proposed a technique where a point is an outlier if in some lower dimensional projection it is present in a local region of abnormally low density. Another clustering based outlier detection technique was proposed in [7] that involves fixed width clustering with w (radius) and is followed by sorting of clusters in the next phase. Harkins et al [10], introduced replicator neural network (RNN) based technique to find outliers.

Lot of work for outlier detection in static data sets has been done but only a small amount of work has been done for outlier detection in dynamic data streams. To overcome these issues, Manzoor [4] described a clustering based approach for outlier detection in dynamic data streams based on k-median approach discussed in [1].

To address the above issues of dynamic data streams, we proposed an algorithm that is a clustering based approach to detect outliers using k-median [1]. It will cluster the data into more than k clusters (facilities) and rather than taking the complete data summary we will only keep the weighted medians from one phase to the next phase making space for the new incoming data points. The medians whose weight does not change after completion of a phase is treated as a temporal candidate outlier for that phase and is checked for the next fixed number of phases to finally declare it as an inlier or a real outlier.

## 3 PROBLEM DEFINITION

In this section, we discuss an algorithm (KORM) to find outliers in data streams which will be better than the already existing methods. Here we formally define data streams, k-median objective and outlier detection over data streams.

Definition 1 (Data Stream): A Data Stream $X = \{x_1, x_2, ..., x_n\}$ possibly infinite series of objects. $x_i$ is represented by n-dimensional vector i.e., $x_i = (x_1^i, x_2^i, ............, x_n^i)$ therefore data projected to the $k^{th}$ dimension from the data stream X is represented as

$$X^k = x_1^k, x_2^k, ............, x_n^k$$

Definition 2 (k-median): Given a set X of n points from some metric space, an integer k, and k members $c_1, .., c_k$ of the metric space, the k–Median cost of using $c_1, ..., c_k$ as medians for X (or, simply, the k–Median cost of $c_1, ..., c_k$ on X) is

$$\sum_{x \in X} \min_{1 \le i \le k} \{dist(x, c_i)\}.$$

So the cost of a set of medians is the value of the k–Median objective function. The k-median objective function is to minimize the sum of assignment distances. Each assignment distance over real spaces is replaced by their squares.

Definition 3 (Outlier): Given a data space in multiple dimensions $A_1, ......, A_m$ with domains $D_1, ......, D_m$ respectively, let the data stream D be a sequence of data objects, where each data object $t \in D_1 X ...... X D_m$. Our task is to online detect if a new coming data is an outlier.

Let data stream objects are elements of a metric space on which we can define a distance function. We consider a data stream as chunks of data

$$X = X_1, X_2, ............, X_m$$

where every chunk contains specified number of n points.

In order to deal with the processing of data stream to find outliers we proposed an algorithm, that would for a given chunk of data find weighted medians (the weight of the median is the number of points assigned to it ) and temporary outliers and delete that summary chunk of data from memory. By this way the freed memory could be used for the next upcoming data chunk. In Fig. 2 below, only the weighted medians marked with black are passed to the next phase and the points allotted to the medians are deleted.

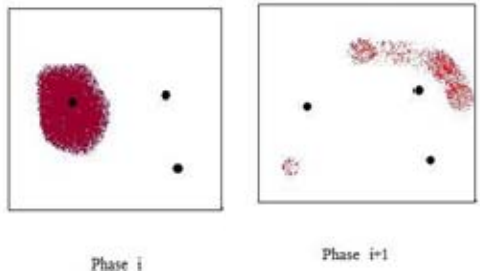

Fig.2. Weighted medians passed from Phase i to Phase i+1

In our method we do not declare a point as an outlier for the data stream but call it a temporal outlier for the given chunk of data. This point may be an outlier for the present data but may not be an outlier for the next data chunk as the data stream is dynamic .We would check this temporal candidate outlier for a given number of data chunks and if it is an outlier for the next given fixed number of stream chunks we declare it as an outlier for the data stream and is not included further for clustering.
Most of the existing work use k-NN approaches for outlier detection which involves parameters R and k, to be provided by the user and require pair wise distance calculations which makes it computationally expensive. The clustering approach for outlier detection using k-mean provides a better solution than k-NN but it is very rigid about the value of k . It provides good clustering results and at the same time deserves good scalability. However , our algorithm for detecting outliers relaxes the value of k and uses minimum k-medians(facilities) and the maximum k log(n) . The flexibility in the value k , assures better stability to our k-median solution than the k-means . It runs in poly-logarithmic space i.e. O(k poly log(n)) and results in a constant factor O(1) approximation ,the algorithm is randomized and has high probability .

## 4 PROPOSED ALGORITHM IN DETAIL

For our proposed algorithm the clustering approach used is k-median approach [1] that relaxes the k-value in the intermediate steps of the algorithm. Each new median ( facility ) is associated with a facility cost and all the assignment distances are associated with assignment cost. The main motive is to minimize the cost of the solution which is sum of assignment cost and facility cost and get good outlier results. In this algorithm we are more concerned for the outliers than the clustering part of it, so we would not as in [1] solve the k-median problem on the final set of O(k log n) medians .

First take the first data point as the first facility. In processing the next point x of weight w, the probability of building a facility at x is min ( w*dist to the closest already open facility ,1).This process of reassignments continues until the cost of the solution and the number of medians does not exceed a specified limit. The weighted medians (facilities) to which no new point has been assigned are treated as temporal candidate outlier for that data chunk. By discarding the safe region (the points that are assigned to some facility), we are able to free memory to accommodate next data chunk. The temporal candidate outliers along with the weighted medians are passed on to the next phase to be processed along with the new chunk of data. The temporal candidate outliers are checked for outlier score for the next fixed number (O) of data chunks. The temporal outlier is declared as an inlier or outlier if its outlier score becomes equal to the user specified limit of O.

THE ALGORITHM : **K-median based OutlieR Miner (KORM)**

*Require n* = number of points in a given data stream ( n = |X| )
*Require k*: the lower bound for the number of facilities ( k < n)
*Require p* : the number of phases and p is assumed to be always less than n ( p ≤ n)
*Require O* : till how many chunks to test outlierness
*Require : γ and β* : constants that should satisfy the condition :
$$\gamma + 4(1+4(\beta+\gamma)) \le \gamma\beta$$
and to make sure that every phase reads at least one more data point we also use a c-approximate algorithm for our approach and so the following condition must also be satisfied
$$\beta \ge 2c(1+\gamma)+\gamma.$$
By experimental evaluation the value of γ and β both is 34.
*Require X* : data stream with $X_1, X_2, \ldots, X_N$ as data chunks.
*Require $X_j$* = { $x_1, x_2, \ldots, x_{Num}$ }. Each point has a weight { $w_1, w_2, \ldots, w_{Num}$ } respectively, associated with it which represents the number of points associated with it. Initially all $w_i$ =1 (1 ≤ i ≤ Num).
*Require Num* : chunk size
*Compute $L_j$* : Each phase has a lower bound value associated with it where $L_{j+1} = \beta L_j$.
*Compute $F_j$*: Each phase has a facility cost associated with it where $F_j = L_j / ( k ( 1+ \log n) )$.

1. $L_1 \leftarrow$ Set_lb( X, k ) / β;
2. i←1; $X_1 \leftarrow$ X ;
3. while there are unread points in X :
   i) $WM_j$ = CLUSTER($X_j, L_j$, k, n ) ;
   ii) If after O number of phases , Outlier Score (Temporal Candidate Outlier) = O ( i.e. the user specified check value ),
      a. Mark it a Real Outlier (outlier $_r$);
      b. $WM_j = WM_j$ –Outlier (outlier $_r$);
      Else
   iii) Mark it as an inlier and remove it from the list of temporal outliers.
   iv) $X_{j+1} = WM_j\ ||\ ( X – R_j )$;

     v) $L_{j+1} = \beta L_j$ ;
     vi) j ← j+1 ;
4. Return the weighted medians $WM_{j-1}$ given by the recent invocation of CLUSTER and also return the points marked recently as real outliers .
5. Exit.

KORM operates in phases and in each phase j , it invokes ONLINE-FL (by Meyerson[12] ) on a modified version $X_j$ of the input stream X and calculates weight for each median $WM_j$ found by ONLINE-FL for $X_j$. Each median found during phase j is then associated with its weight and the weighted set of medians is the " the solution from phase j " to the next phase.
     The algorithm KORM in step 3(i) , by making call to CLUSTER returns all the weighted medians including the temporal candidate outliers that it found while processing data chunk $X_j$, that has been read. In step 3(ii), we check for the outlier score of temporal outliers for o number of phases, to declare them as an inlier or a real outlier. In step 3(ii) (b) we exclude the real outliers from the weighted medians ,so that it is not again checked by new data chunk for clustering . In step 3(v), we concatenate weighted medians from previous chunk ( by removing the rest of data points in previous chunk to make space for new data ) and X- $R_j$ ( the portion of X that is still unread after the end of phase j) .

The first call to KORM sets the lower bound using the following algorithm :

**Set_lb ( data stream X , integer k )**
1. Let D equal the distance between the closest pair of the first k+1 members of X in order;
2. Return D;

**CLUSTER (data stream $X_j$ , lower bound $L_j$ , int k, int n)**
1. Run 2 log n parallel invocations of ONLINE-FL with facility cost $F_j$ on the input data chunk $X_j$ .
2. Each invocation is run as long as the cost of the solution on the modified input does not become greater than $4L_j (1+ 4 (\gamma+\beta))$ or the number of medians does not exceed $4k (1+\log n)(1+ 4 (\gamma+\beta))$ or the cost of the solution remains same for two or more invocations.
3. Continue until all invocations have been stopped
4. Mark as "read "all the points seen by this invocation except the last point (that may exceed the cost limit or medians limit).
5. Mark the medians as temporal candidate outliers (TCO ) if there is no increase in its weight in this phase and increase its outlier score by one .
6. Return the medians (including the temporal candidate outliers) found by this invocation.

In step 4 of CLUSTER we mark all the points that have been read in a given phase except the last point. The notion of read points is to obtain a bound on the number of phases by ensuring that every phase makes progress by reading at least one new point. When we change phases, our algorithm might see a point but not mark it as read so we keep the last point as unread to be processed again in the next phase.

Here is a brief overview of ONLINE-FL by A.Meyerson [12] .Given a point stream ONLINE-FL ,we can find a set of medians for an initial segment of that stream .

**ONLINE-FL ( data stream S, facility cost f )**
1. Make one pass over X performing the following steps for each weighted point x ∈ X .
2. Let θ be the distance of the current point x (with weight w ) to the closest already-open median.
3. The probability of opening a new median/facility at x is min ( θw/ f , 1) , otherwise the new point is assigned to the closest already-open facility.

## 5 ALGORITHM DESCRIPTION

X is the data stream consisting of $X_1, X_2, \ldots, X_N$ data chunks where every chunk contains fixed number Num of data points.
     Let $WM_j$ denote the group of weighted medians of the $j^{th}$ chunk of stream. $C_j = C_{1j}, C_{2j}, C_{3j}, \ldots, C_{Lj}$ where $k \leq L \leq 4k (1+\log n) (1+4 (\gamma + \beta))$ and $1 \leq j \leq N$ . We here have a range of permissible value of k which is the advantage given by k-median of the flexibility in the value of k .Every $WM_j$ is checked for outlierness .
     In the proposed algorithm the first step is input chunks of data one after the other, $X_j =\{ x_1, x_2, x_3, \ldots, x_{Num} \}$ ,where $X_j$ represents $j^{th}$ chunk while $x_i$ is the $i^{th}$ element in the chunk $X_j$. Procedure for outlier detection using k-median is simple and we start by considering the first element in the first data chunk as the assumed median ( facility ). Every next data point is then considered for either being a facility or being assigned to already existing closest facility as described in Step 3 of ONLINE-FL.
     When a point is assigned to an existing facility, the weight of the facility is increased by the weight of the point and the totalcost of the solution as well as the cost of the point is increased by the weight of point* distance from the closest median .When a point is itself declared as a new facility , the totalcost of the solution is increased by facility cost for that phase.
     After all the assignments of the data points in the given data chunk , reassignments are done until the number of medians (facilities) does not exceed $4k (1+\log n) (1+4(\gamma + \beta))$ or the cost of solution does not exceed $4L_i (1+4(\gamma + \beta))$ . At the end of this phase, we remove the safe points and check the weighted medians for outlierness. If no point is assigned to the median, that facility (median) is declared as a temporal candidate outlier for that data chunk and its outlier score is increased by one. After O number of phases (from the phase it has been declared as temporal candidate outlier ),we check its outlier score . If it is O, we declare the median as real outlier and further not use it for clustering otherwise if outlier score after O number of phases is less than O ,we consider it as an inlier . All the weighted medians in each phase are passed as input to the next phase and the process is again repeated with a new value of facility cost and lower bound.
     The value of k in our algorithm determines the quality of our result and the value O has to be selected carefully, by keeping in view the value of parameter k and Num .Our algorithm is found to be more stable than k-means

[4] , as it relaxes the value of k and provides a solution that uses poly-logarithmic space .

## 6 EXPERIMENTAL EVALUATION

Experiments were conducted to justify that our algorithm and get good results. We conducted all experiments on a Windows Vista Home Premium with Intel® Core™ Duo CPU T2450 @ 2.00 GHz with 1.00 GB RAM. Experiments were conducted in Matlab 7.0.4 on various data sets using gamma and beta value to be 34( found by experimental evaluation). Here we are more interested in the outliers rather than the clustering quality of the results. We used datasets from the UCI Machine Learning Repositoiry [13].

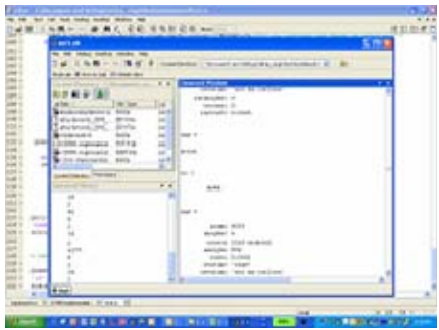

Fig. 3. KORM on Abalone Dataset in Matlab

Experiments were conducted on the mixed attribute value Abalone data set to predict the age of abalone and the multivariate Teaching Assistant Evaluation dataset to classify their teaching performance into three classes: low, medium or high. Fig. 3 shows the Matlab editor and the command window while implementing KORM on Abalone Dataset. Table 1 shows the results of KORM implementation on the the Abalone dataset.

TABLE 1
DATA RESULTS OF KORM ON ABALONE DATASET

| Data set name | | Abalone Dataset |
|---|---|---|
| Type | | Mixed attribute value |
| No. of instances | | 4177 |
| No. of attributes | | 8 |
| INPUTS | K | 2 |
| | O(CountValue) | 2 |
| | Gamma ( γ ) | 34 |
| | Beta ( β ) | 34 |
| OUTPUTS | No. of Outliers | 4 |
| | Sum of Squared Distance | 4.6003e+004 |
| | Cost of solution | 3.4445e+005 |

*The value of e=10*

The results in Table 1 shows that the total sum of distances value has been minimized as against the largest assignment distance in k-means [4] making it more acceptable solution for outlier detection.

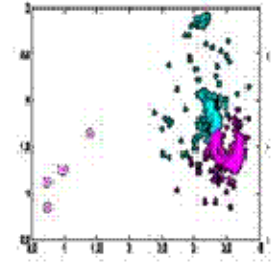

Fig. 4. Outliers in sparse region in Abalone Dataset

We compared our results with DK (distance based k-nearest neighbor) approach. This method based on distance of a point from its $k^{th}$ nearest neighbor. It ranks points on the basis of distance to its $k^{th}$ nearest neighbor and declare the top-n points in the ranking as outlier. We also compared our approach with the CORM method based on k-means [4]. Fig. 4 shows the graphical results after implementation of KORM on Abalone dataset giving four outliers marked as pink that are present in the sparse region . The graphical results were similar to that of DK approach as well as k-means approach but our approach was better in terms of stability, computational time and space.

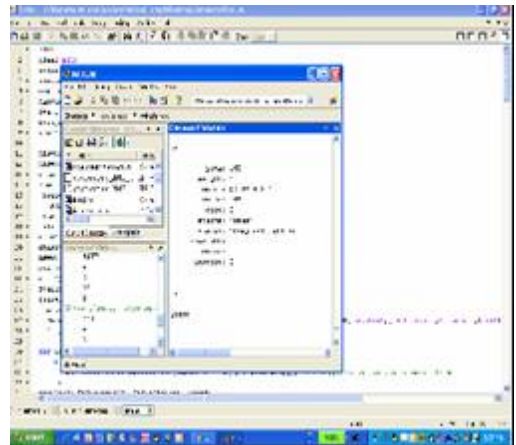

Fig. 5. KORM on Teaching Assistant Evaluation Dataset in Matlab

Fig. 5 shows the Matlab editor and the command window while implementing KORM on Teaching Assistant Evaluation Dataset. Table 2 shows the results of KORM implementation on the the Teaching Assistant Evaluation dataset.

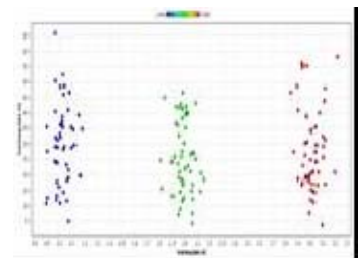

Fig. 6. No Outliers in Teaching Assistant Evaluation Dataset

TABLE 2
DATA RESULTS OF KORM ON TEACHING ASSISTANT EVALUATION DATASET

| Data set name | | Teaching Assistant Evaluation |
|---|---|---|
| Type | | Multivariate |
| No. of instances | | 151 |
| No. of attributes | | 5 |
| INPUTS | K | 3 |
| | O(CountValue | 3 |
| | Gamma ( γ ) | 34 |
| | Beta ( β ) | 34 |
| OUTPUTS | No. of Outliers | Nil |
| | Sum of Squared Distance | 44319 |
| | Cost of solution | 1.7941e+005 |

Fig. 6 shows the graphical results of implementation of our algorithm on multivariate Teaching Assistant Evaluation dataset to evaluate teacher's performance. It shows the three clusters having almost equal number of elements and no outliers depicting that shows there are equal number of teachers in the three scores: low, medium and high. Some interesting results of our experiments using proposed method as compared to the Dk and the k-means approach are shown in Table 3.

In case of DK-Outliers to calculate CPU time, the values used are K = 3; n = 7 where n = number of values we want to identify while K = nearest neighbors to consider in the calculus of Dk, while for Dk-Outliers Nested Loop D = 0.45; p = 0.95.

TABLE 3
CPU TIME IN SECONDS

| N | Proposed | k-means | DK-Outliers | Dk-Outliers NL |
|---|---|---|---|---|
| 4177 | 0.212134 | 0.212416 | 7.906283 | 88.814268 |
| 151 | 0.011522 | 0.011121 | 0.261252 | 2.737831 |

The experimental observations as shown in Table 2 show that our approach is comparable to the k-means approach but provides us with a better solution in terms of stability and space utilization as it uses only poly-logarithmic space i.e. O ($k\ poly\ log\ n$) .

## 7  CONCLUSIONS

In this paper, we present a cluster-based outlier detection approach based on k-median .In our method, we give more attention to the outliers than to the clustering part of our result. As for data stream , more points keep on flowing in, so we discard the safe region to free the physical memory for the next generation of data to be processed , effectively. We give the candidate outliers a chance of survival in the next incoming chunks rather than declaring them real outliers in the current chunk, to cover for the dynamics of the data streams. Our approach provides a more stable solution than the k-mean approach by Manzoor [4] as it confines the parameter k to a range of values rather than a fixed value .Our k-median approach gives us a better clustering solution and better defined outliers than the k-mean solution as it tries to minimize the sum of assignment distances, the total cost and space. The implementation of our algorithm on various datasets shows that successful results are obtained using same value for k and O. For future work, we need to improve our approach to make it more time and space efficient. Our approach only deals with numerical data so in our datasets all the character data had to be first converted to its Ascii values , so future work requires modifications that can make it applicable for textual mining also. The approach needs to be implemented on more complex datasets using digital processors.

## REFERENCES

[1] Moses Charikar, Liadan O´Callaghan, Rina Panigrahy, Better Streaming Algorithms for Clustering Problems. In *Proc. Of STOC'03, June 9–11, 2003, San Diego, California, USA. 2003 ACM*

[2] Ramaswamy S., Rastogi R., Kyuseok S.:Efficient Algorithms for Mining Outliers from Large Data Sets,*Proc. ACM SIDMOD Int. Conf. on Management of Data, 2000*

[3] M.M. Breunig, H.P.Kriegel, R.T. Ng and J.Sander, LOF: Identifying Density-Based Local Outliers *ACM SIGMOD 2000*

[4] Manzoor Elahi, Kun Li , Wasif Nisar, Xinjie Lv ,Hongan Wang , Efficient Clustering-Based Outlier Detection Algorithm for Dynamic Data Stream.In *Proc. Of the Fifth International Conference on Fuzzy Systems and Knowledge Discovery(FSKD.2008)*.

[5] Angiulli, F. and Fassetti, F. Detecting Distance-based Outliers in Streams of Data. In *Proc. of the Sixteenth ACM Conf. on information and Knowledge Management (Lisbon, Portugal, November 2007).CIKM '07*.

[6] Charu C. Aggarwal, Philip S. Yu, Outlier Detection for High Dimensional Data, *Proc. of the 2001 ACM SIGMOD int. conf. on Management of data*, p.37-46, May 21-24, 2001, Santa Barbara, California, United states

[7] E. Eskin, A. Arnold, M. Prerau, L. Portnoy, and S. Stolfo. A Geometric Framework for Unsupervised Anomaly Detection: Detecting Intrusions in Unlabeled Data. *In Data Mining for Security Applications, 2002.*

[8] Knorr, E. M., Ng, R.T. Algorithms for Mining Distance-Based Outliers in Large Datasets, *Proc. 24th VLDB, 1998*

[9] A. Arning, R. Agrawal, P. Raghavan. A Linear Method for Deviation Detection in Large Databases. *In: Proc of KDD'96, 1996:* 164 -169.

[10] S.Harkins, H.He, G. J.Willams, R.A. Baster. Outlier Detection Using Replicator Neural Networks. *In: Proc of DaWaK'02, 2002:* 170-180.

[11] Dragoljub Pokrajac ,Aleksandar Lazarevic ,Longin Jan Latecki . Incremental Local Outlier Detection for Data Streams . *IEEE Symposium on Computational Intelligence and Data Mining (CIDM)*, April 2007

[12]  A.Meyerson. Online Facility Location. *Proc. FOCS, 2001.*
[13]  Asuncion, A.and Newman,D.J. "2007".UCI Machine Learning Repository Irvine,CA University of California, School of Information and Computer

**Parneeta Sidhu ,**  B.Tech-CSE(2002), M.tech-IS(2009), Member of the IEEE. Presently working as Senior Lecturer in KIIT College of Engineering. Published many papers in National  and International- Conferences in the field of Data Mining and Knowledge Engineering.

**MPS Bhatia , B.Tech,M.tech and PhD in Computer Science.** Presently working as Professor in NSIT,  New Delhi. Under my guidance many students have submitted their M.Tech  Thesis and  successfully completed their PhD.

**Priti  Bansal** , B.Tech-CSE(2001), M.tech-IS(2009). Presently working as Senior Lecturer in Ideal Institute of Technology, Ghaziabad. Published many papers in National  and International Conferences in the field of Data Mining and Knowledge Engineering.